\documentclass[oldversion]{aa} 
\usepackage{natbib}
\usepackage{graphicx}
\usepackage{amsmath}
\usepackage{txfonts}
\usepackage{footmisc}
\begin{document}
   \title{The {\it Chandra} X-ray view of the power sources in Cepheus A}

   \author{P. C. Schneider
          \and
          H. M. G\"unther
          \and
          J. H. M. M. Schmitt
          }

   \institute{Hamburger Sternwarte,
              Gojenbergsweg 112, 21029 Hamburg\\
              \email{cschneider/mguenther/jschmitt@hs.uni-hamburg.de}
             }

   \date{Received 28.07.2009 / accepted 02.09.2009}

  \abstract
   {
   The central part of the massive star-forming region Cepheus A contains several radio sources which indicate multiple outflow phenomena, yet the driving sources of the individual outflows have not been identified. We present a high-resolution \textit{Chandra} observation of this region that shows the presence of bright X-ray sources with luminosities of $L_X \gtrsim 10^{30}$~erg$\,$s$^{-1}$, consistent with active pre-main sequence stars, while the strong absorption hampers the detection of less luminous objects. A new source has been discovered located on the line connecting H$_2$ emission regions at the eastern and western parts of Cepheus~A. This source could be the driving source of HH~168. We present a scenario relating the observed X-ray and radio emission.
   }

   \maketitle
%


\section{Introduction\label{sect:intro}}
Stars form in collapsing molecular clouds. The largest clouds harbour very massive protostars in addition to the numerous late-type stars.
\object{Cepheus~A} is the second nearest high-mass star forming region at a distance of about 730~pc \citep{Johnson_1957}. 

The Cep~A region contains multiple outflows, e.g. large scale molecular outflows, extending several arcmin ($\sim 10^5$~AU) from their driving source with velocities of a few 10 km$\,$s$^{-1}$. The complicated outflow geometry has been explained by dense condensations, which redirect the outflow \citep[][]{Codella_2003,Hiriart_2004}, but can be also interpreted as independent outflows or by the evolution of a single outflow.

\begin{figure}
  \centering
   \includegraphics[width=0.49\textwidth]{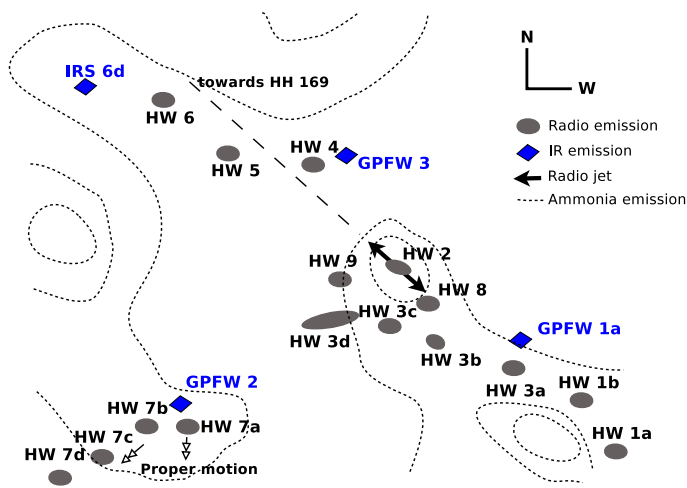}
   \caption{Sketch of some emission sources in the Cep~A region. See Fig.~\ref{fig:CenterXray} for the exact location of the individual sources. The ammonia emission is sketched from the observations of \citet{Torrelles_1993}. }\label{fig:CenterOverview}%
\end{figure}

The line of sight towards the centre of Cep~A suffers from strong absorption and hampers optical observations of the central sources. From reflected light measurements in the \mbox{infrared (IR)}, \citet{Lenzen_1984} derived $A_V \gtrsim 75$ while \citet{Goetz_1998} estimated $A_V \gtrsim 200$ from their inability to detect the central radio source in their IR-images (assuming a B0-type star).
\citet{Sonnentrucker_2008} derive the hydrogen absorbing column density from the silicate absorption feature at 9.7~$\mu$m to range from a few $10^{22}$~cm$^{-2}$ to higher than $10^{23}$~cm$^{-2}$.
Therefore, only radio observations provide detailed maps of this region.
Several distinct sources with different apparent sizes and spectral indices have been revealed by cm-wavelength observations \citep[][]{Hughes_1984, Garay_1996}; we also use their nomenclature ``HW1\dots9'' to designate radio sources in that region (as sketched in Fig.~\ref{fig:CenterOverview}).

The most radio luminous object of these HW-sources is HW~2, which probably harbours a B0.5~--~B2 star with a mass of about 15~--~20 M$_{\sun}$ \citep{Hughes_1984, Garay_1996, Curiel_2006}. This source is usually refered to as the center of the Cep~A region. A thermal radio-jet with a position angle of $\sim 45\,^{\circ}$ is present \citep{Curiel_2006}, indicated by arrows in Fig~\ref{fig:CenterOverview}. 
It is probably related to (at least) one part of the large scale molecular outflow, which has a similar position angle and is directed towards HW~2.
The jet contains knots moving with velocities of about 500~km$\,$s$^{-1}$ as derived from high resolution cm-wavelength observations \citep{Curiel_2006}.
Some of the radio sources, located approximately in the direction of this jet at distances of a few 10~arcsec (HW~1, HW~4, HW~5 and HW~6), are likely shock-induced free-free emission components (with non-thermal contributions) powered by the HW~2 jet \citep{Garay_1996}.
There has been some debate on the interpretation of the observations very close to HW~2 itself. Some observations indicate the presence of a disk \citep[e.g. HCO$^+$, SiO, CH$_3$CN, SO$_2$, 335~GHz continuum;][respectively]{Torrelles_1996,Gomez_1999,Patel_2005, Jimenez-Serra_2007, Torrelles_2007}, while signatures of radio emission explained by the presence of young stellar objects (YSOs) have also been found \citep{Curiel_2002, Brogan_2007, Comito_2007}. \citet{Jimenez-Serra_2007} show that disk signatures are present very close to HW~2, suggesting a disk with a size of about 600~AU.

The radio complex HW~3, located 3~--~4~arcsec south of HW~2 (see Fig.~\ref{fig:CenterOverview}), constitutes an elongated structure oriented primarily in the east-west direction, which is resolved at higher resolutions.
Due to their spectral properties and dense cores or association with masers, all subcomponents have been proposed to harbour an internal energy source; i.e., they could be associated with an YSO \citep[e.g.][]{Hughes_2001, Garay_1996, Brogan_2007}. However, there is no consensus about their nature in the literature.
In the vicinity of HW~3a, 3b and 3d water maser activity has been found \citep{Cohen_1984, Torrelles_1998}, while HW~3c has an associated submillimetre core \citep{Brogan_2007} and a composite spectrum, suggesting the presence of a jet \citep{Hughes_2001}.
HW~3a breaks up into two time-variable components; one might be associated with an infrared source \citep[GPFW~1a (diamond in Fig.~\ref{fig:CenterOverview}):][]{Goetz_1998, Lenzen_1988, Garay_1996, Hughes_1997}.
HW~3d also breaks up into at least four distinct objects in high-resolution observations.

HW~8 and HW~9, located close to HW~2 and HW~3 (a few arcsec distance), are two relatively compact radio sources, showing large variations in their flux densities and are therefore assumed to be associated with low-mass pre-main sequence stars \citep{Hughes_2001, Garay_1996}.

HW~7 consists of several radio emission regions about 20\arcsec~ south-east of HW~2. The main components b, c and d are aligned at a position angle of 107$^\circ$, pointing towards just south of HW~2 at one of the HW~3 sources.
Therefore, \citet{Garay_1996} proposed that HW~3d drives an outflow directed towards HW~7, which is interpreted as shock induced radio emission. From their proper motion measurements of the individual components of HW~7, \citet{Curiel_2006}  propose that the driving source is located just north of the HW~7a component, mainly because HW~7a moves almost perpendicularly to the rest of the HW~7 sources due south. We identify this source with GPFW~2 \citep{Goetz_1998}.

Just as radio waves, X-rays penetrate through a high absorbing column density and are therefore a useful tool to disentangle individual emission components.
In the case of Cep~A, B-type stars are potentially associated with the radio sources. They have a typical X-ray luminosity of \mbox{$\sim10^{30}$~erg$\,$s$^{-1}$} for late B stars and up to \mbox{$\sim10^{32}$~erg$\,$s$^{-1}$} for B0 stars \citep{Berghoefer_1997}. The $L_X / L_{bol}$ ratio increases from $10^{-7}$ for mid-B stars to higher values for A stars and later objects. A-type stars have X-ray luminosities of usually just below $10^{30}$~erg$\,$s$^{-1}$ and later stars, even young M-dwarfs, can easily emit more than $10^{29}$~erg$\,$s$^{-1}$ \citep{Preibisch_2005}.
Typical mean plasma temperatures for later-type stars are about 1.4~keV, while early-type stars can have relatively low temperatures of only a few 0.1~keV. A study of young O and B-type stars in the Orion nebula \citep{Stelzer_2005} showed that a second temperature component around 2~keV is mostly present in the early and mid B-type stars.

Although column densities above $10^{23}$~cm$^{-2}$ almost completely absorb any soft stellar X-ray emission ($E < 1$~keV),  higher energy photons remain detectable in the {\it Chandra} observation.
Cep~A was already observed by {\it XMM}-Newton \citep{Pravdo_2005}. That observation found hard X-ray emission from the region around HW~2, but was not able resolve the individual sources due to the large point-spread function (PSF) of {\it XMM}-Newton; the authors speculate that deeply embedded protostars or high-velocity outflows might be responsible for the observed X-ray emission.

The focus of this article is to use {\it Chandra}'s high angular resolution to disentangle the sources in the central region of Cep~A. 

\section{Observation, data processing and analysis}
\subsection{Observation}
\textit{Chandra} observed Cep A on 2008-04-08 for 80~ks with ACIS-I (Obs-ID 8898). The analysis was carried out using CIAO~4.1.2 and the science threads published at CIAO website\footnote{\texttt{http://cxc.harvard.edu/ciao/}}.
The data was not reprocessed, i.e., standard parameters like pixel randomisation are applied to the data. 
\subsection{Processing\label{sect:proc}}
For point source analysis, the extraction regions were chosen to contain about 90~\% of the source photons (1.5~arcsec circles).
The energy range for our analysis is 0.3~keV to 9.0~keV; we therefore define a low band from 0.3~keV to 1.5~keV and a high band from 1.5~keV to 9.0~keV.
We used a background region which contains 72 photons in a circle of 11~arcsec radius (48 photons in the high band). The estimated background within the extraction regions of the individual sources is therefore about 1.3~counts in total, 0.9~photons of them in the high band.
To derive the statistical significance of a source at a specific position, i.e., the probability to find a source at a given position by chance, we run a source detection algorithm (\texttt{celldetect}) on the central field of view (three~arcmin radius), where the PSF is still narrow. In this region 34 (25) sources with a S/N value above 3(4) are present (not to be mistaken as the significance of the source\footnote{\texttt{http://cxc.harvard.edu/ciao/download/\dots\\ \hspace*{3.5cm}\dots doc/detect\_manual/index.html}}).
The source with the lowest S/N discussed in the following is found by the source detection algorithm at \mbox{S/N = 1.7}. Within the 3~arcmin radius 68 sources were then found. Corrected for the chip gaps,  statistically fewer than 0.01 sources are expected within the extraction region at any position specified a priori.

XSPEC v12.3.1x 
\citep{XSPEC} was used to estimate the plasma  properties of the individual sources. For the spectra an absorbed plasma emission model \citep[APEC,][]{APEC} was chosen.
Errors denote 1$\sigma$ confidence ranges and quoted X-ray luminosities are dereddened luminosities in the \mbox{0.3 -- 10.0~keV} band.

\subsection{Data analysis}
\label{sect:centralregion}
\begin{figure}
  \centering
   \includegraphics[width=0.49\textwidth]{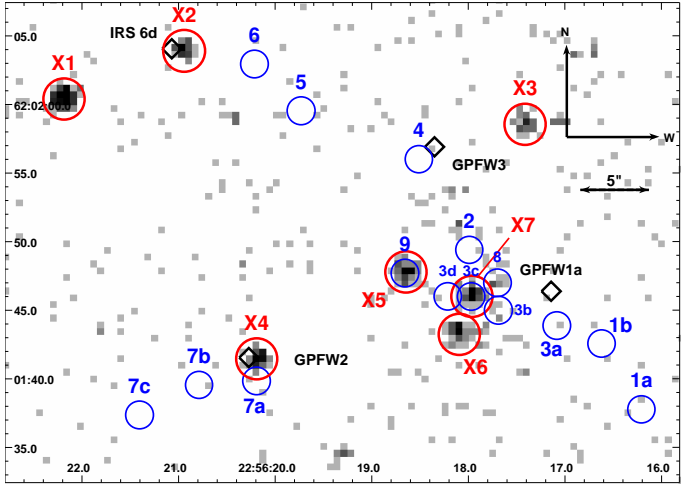}
   \caption{X-ray image (1.5-9~keV) of the central region of Cep A. The radii of the blue circles designating the radio sources are 1.0~arcsec while that of the red circles naming the X-ray sources are 1.5~arcsec. Numbers indicate HW~radio sources. The blacks diamonds are selected infrared sources.}\label{fig:CenterXray}%
\end{figure}

In Fig.~\ref{fig:CenterXray} an overview of the central region of Cep~A in the high band is shown. All X-ray sources are visible; no source in this region is present only in the low band. The locations of radio sources and infrared point-like sources are also indicated (cf. sketch in Fig.~\ref{fig:CenterOverview}).
We summarise the properties of the X-ray sources (detected with a S/N ratio above 3 by \texttt{celldetect}) in Tab.~\ref{tab:X-ray}, where cross-IDs, position, plasma parameters, radio fluxes and infrared luminosities for the detected X-ray sources and limits for the non-detections are given.
Infrared magnitudes were taken from \citet{Lenzen_1984} for the IRS sources and from \citet{Goetz_1998} for the GPFW sources; radio fluxes are from \citet{Garay_1996}. The source position is either given by the location of the respective radio source or by the detection algorithm. The source positions were slightly ($<0.4$~arcsec) adjusted to contain a larger number of counts for X5 and X6 (only towards the south).
The large extinction towards the centre of Cep~A \citep{Sonnentrucker_2008} absorbs the low-energy X-ray emission from embedded stars and only the high band remains observable. A plasma with an equilibrium electron temperature of 1~keV still radiates 13~\% of its energy-loss (3~\% of the photons) in that band. A source with a large fraction of low energy photons would be a foreground object,  while an object with only high energy photons is either located within the cloud or a background object.

    \begin{table*}[ht!]
    \begin{minipage}[h]{0.99\textwidth}
\setlength\tabcolsep{2pt}
    \renewcommand\footnoterule{}

      \caption{X-ray properties of X-ray, radio and infrared sources in the region around HW~2/3. \label{tab:X-ray}}
      \begin{tabular}{c c c c c r c c c c c r r r} \hline \hline
        Source & Radio & Infrared &RA & Dec & Counts\footnote{Including a background of 1.3 cts.}   & Median  & Temperature & Absorption & Luminosity & Radio & Spectral & Infrared \\ 
            ID  & ID & ID       &    &     &          & energy\footnote{in keV} & $kT$~(keV) & $10^{22}$~cm$^{-2}$ & log $L_X$ (erg$\,$s$^{-1}$) & flux (mJy) & index & mag\footnote{K-band}\\ 
        \hline
        X1 & & IRS 6b & 22:56:22.2 & +62:02:00.5 & 137 & 2.3 & 2.0 -- 3.0 & 1.4 -- 2.2 & 30.7 -- 30.9 & \multicolumn{2}{c}{--} & 8.4 \\
        X2 & & IRS 6d & 22:56:20.9 & +62:02:04.0 & 44 & 4.2 & 2.0\footref{fn:fixed} &  10 -- 18 & 30.9 -- 31.1 & \multicolumn{2}{c}{--} & 11.1\\
           & &        &            &             &    &     & 1.0  \footref{fn:fixed} & 20 -- 32 & 32.0 -- 32.3 \\
        X3 & & ?\footnote{visible in the \citet{Goetz_1998} IR-images}& 22:56:17.4 & +62:01:58.6 & 36 & 4.3 & 2.0 \footnote{fixed\label{fn:fixed}} & 10 -- 18 & 30.9 -- 31.1 & \multicolumn{2}{c}{--} \\
           & &                                  &            &             &    &     & 1.0 \footref{fn:fixed} & 17 -- 26 & 31.8 -- 32.1 \\
        X4 & & GPFW~2 & 22:56:20.2 & +62:01:41.4 & 64 & 3.9 & 0.9 -- 1.7 & 11 -- 32 & 31.2 -- 32.3 & \multicolumn{2}{c}{--} & 15.6\\
        X5 & 9  & & 22:56:18.6 & +62:01:47.8 & 60 & 4.1 & 0.7 -- 1.3 &  20 -- 38 & 31.8 -- 33.4 & 3.0 & -0.2\\
        X6 & &  & 22:56:18.1 & +62:01:43.2 & 50 & 5.5 & 2.0\footref{fn:fixed} & 32 -- 50 & 31.5 -- 31.8 & \multicolumn{2}{c}{--} \\
           & &                                  &            &             &    &     & 1.0 \footref{fn:fixed} & 61 -- 83 & 33.1 -- 33.4 \\
        X7 & 3c &  & 22:56:18.0 & +62:01:46.0 & 60 & 6.4 & 2.0\footref{fn:fixed} & 62 -- 106 & 32.1 -- 32.5 & 3.7 & 0.4\\
        & &                                  &            &             &    &     & (1.0 \footref{fn:fixed} & 127 -- 175 & 34.0 -- 34.5)\footnote{This value is unlikely high for a stellar source, see Sect.~\ref{sect:X7}.} \\
           & 1a &  & 22:56:16.2 & +62:01:37.8 &  2 & \multicolumn{3}{c}{--} & $<$29.8 (30.5) & 2.2 & -0.6 \\
	   & 1b &  & 22:56:16.6 & +62:01:42.6 &  2 & \multicolumn{3}{c}{--} & $<$29.8 (30.5) & 2.9 & -0.3 \\
	   &  2 &  & 22:56:19.0 & +62:01:49.4 &  $<$4 & \multicolumn{3}{c}{--} & $<$30.0 (30.7) & 7.5 & 0.7 \\
	   & 3a & GPFW~1a? & 22:56:17.1 & +62:01:43.9 &  1 &  \multicolumn{3}{c}{--} &$<$ 29.6 (30.3) & 0.2 & -- & 14.7\\
	   & 3b &  & 22:56:17.7 & +62:01:45.0 &  6 &  4.6 & \multicolumn{2}{c}{\hspace{-31pt}--} & 29.9 (30.6)  & 5.0 & 0.0 \\
	   & 3d &  & 22:56:18.2 & +62:01:46.0 &  3 &  \multicolumn{3}{c}{--} &$<$29.9 (30.6) & 7.7 & 0.3 \\
	   & 4  & GPFW~3 & 22:56:18.5 & +62:01:56.0 &  1 &  \multicolumn{3}{c}{--} &$<$29.6 (30.3) & 4.4 & -0.2 & 12.6 & \\
	   & 5  &  & 22:56:19.7 & +62:01:59.5 &  1 &  \multicolumn{3}{c}{--} &$<$29.6 (30.3) & 1.3 & -0.5 \\
	   & 6  &  & 22:56:20.2 & +62:02:03.0 &  0 &  \multicolumn{3}{c}{--} &$<$29.2 (29.9) & 3.8 & -0.3 \\
	   & 7a &  & 22:56:20.2 & +62:01:39.9 &  2 &  \multicolumn{3}{c}{--} &$<$29.8 (30.5) & 9.6 & -0.1 \\
	   & 7b &  & 22:56:20.8 & +62:01:39.6 &  1 &  \multicolumn{3}{c}{--} &$<$29.6 (30.3) & 2.8 & -0.4 \\
	   & 7c &  & 22:56:21.4 & +62:01:37.4 &  0 &  \multicolumn{3}{c}{--} &$<$29.2 (29.9) & 3.8 & -0.4 \\
	   & 7d\footnote{Data from \citet{Hughes_1984}} &  & 22:56:21.9 & +62:01:36.5 &  1 &  \multicolumn{3}{c}{--} &$<$29.7 (30.4) & 2.4 & -- \\
	   & 8  &  & 22:56:17.7 & +62:01:47.0 &  9 & 6.2 &  \multicolumn{2}{c}{\hspace{-31pt}--} &30.1 (30.8) & $<$0.12 & -- \\
	   &  & IRS 6c & 22:56:17.5 & +62:01:51.1 & 1 & \multicolumn{3}{c}{--} &$<$29.6 (30.3) & \multicolumn{2}{c}{--} & 8.5\footnote{L-band}\\
        \hline
      \end{tabular}
    \end{minipage}
    \end{table*}

\begin{figure}
  \centering
   \includegraphics[width=0.24\textwidth]{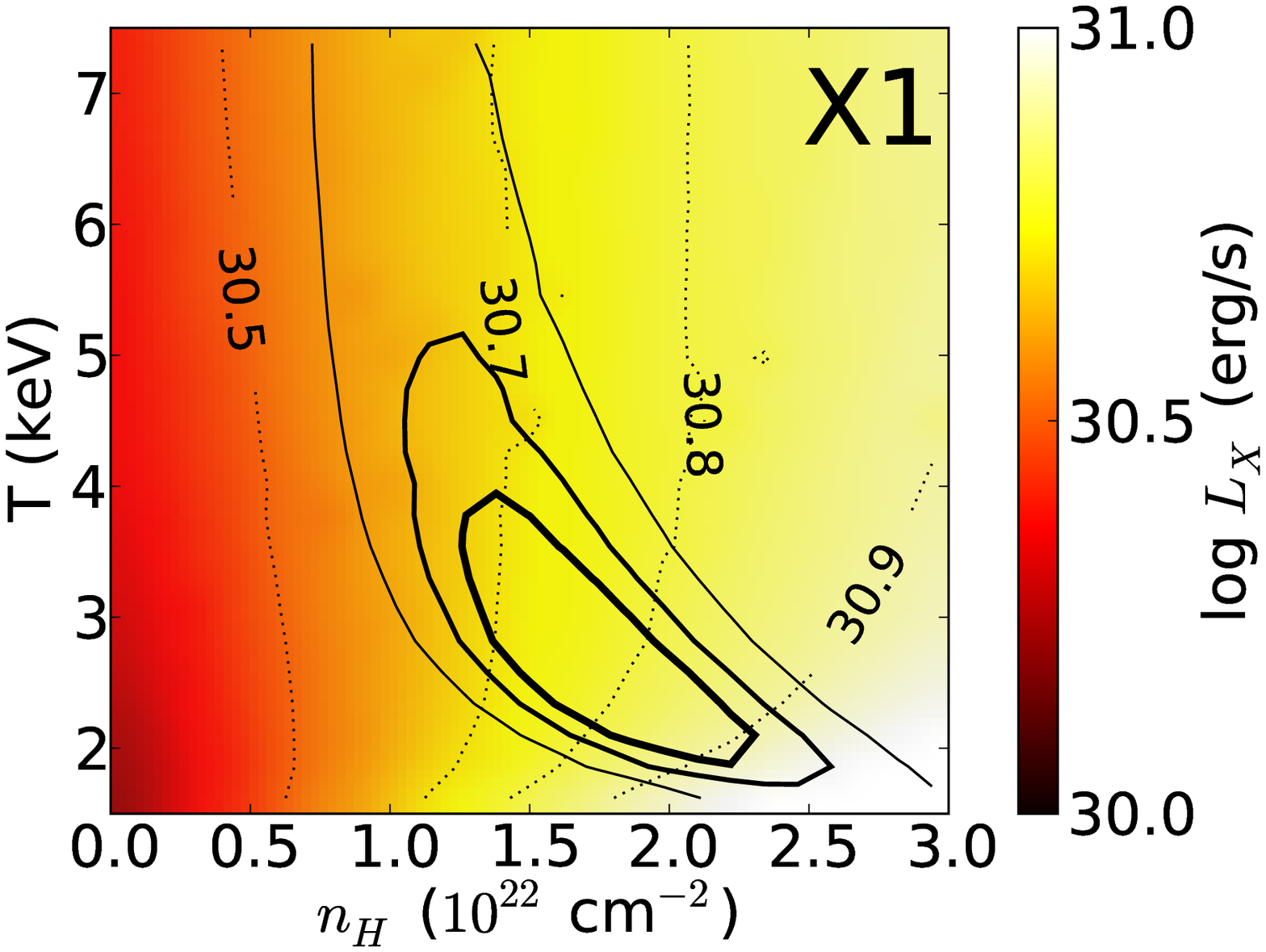}
   \includegraphics[width=0.24\textwidth]{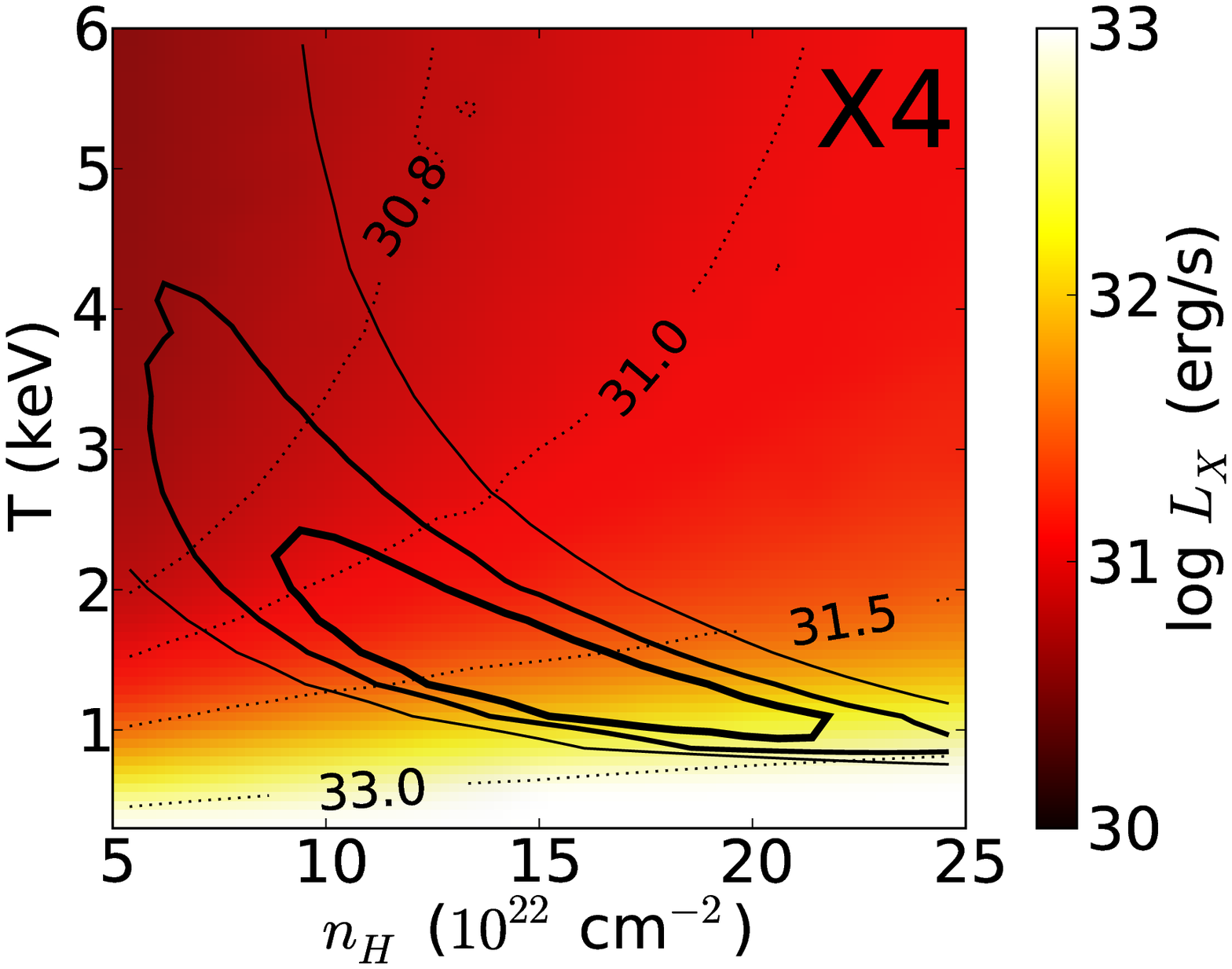}
   \includegraphics[width=0.24\textwidth]{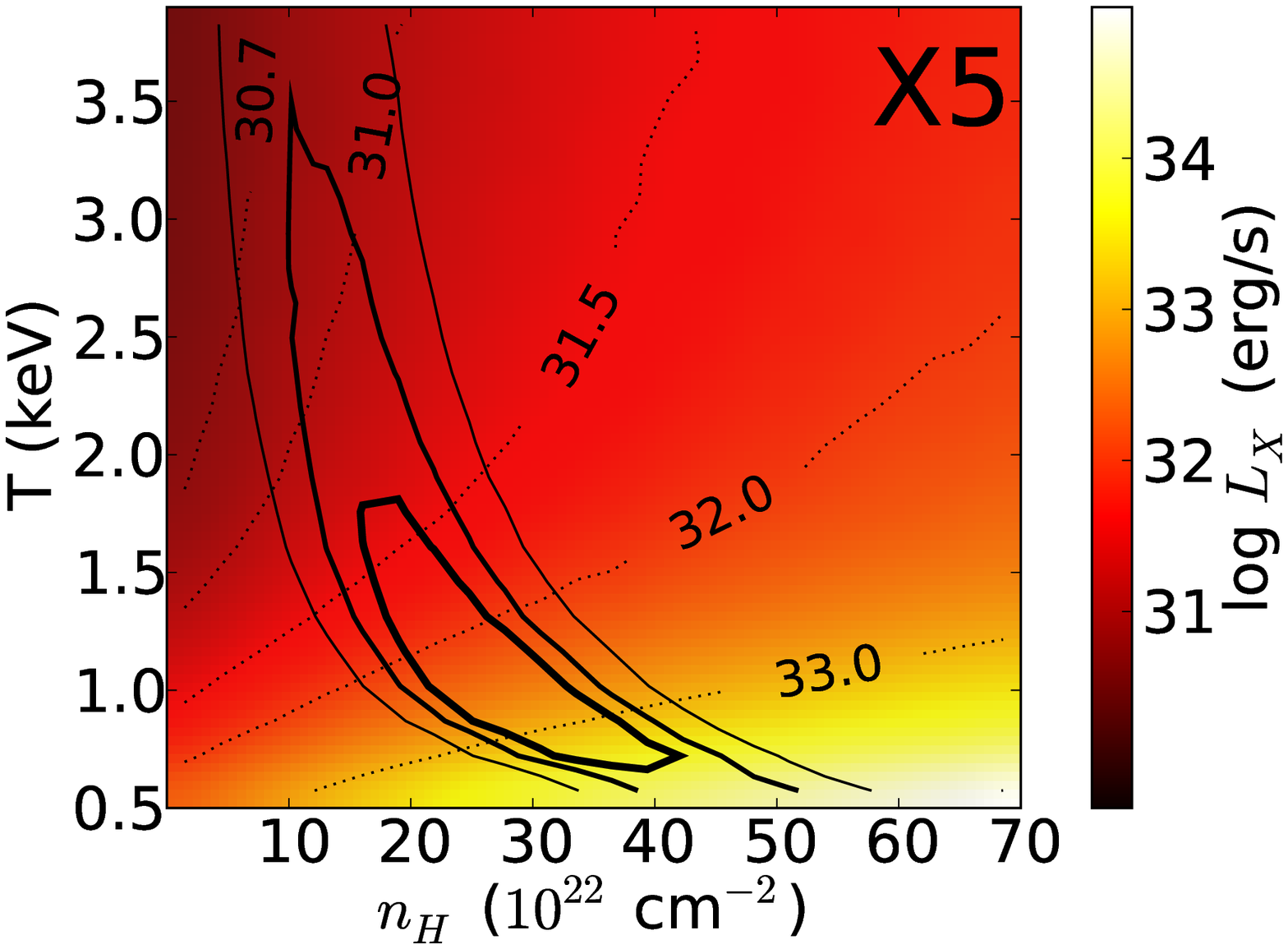}
   \includegraphics[width=0.24\textwidth]{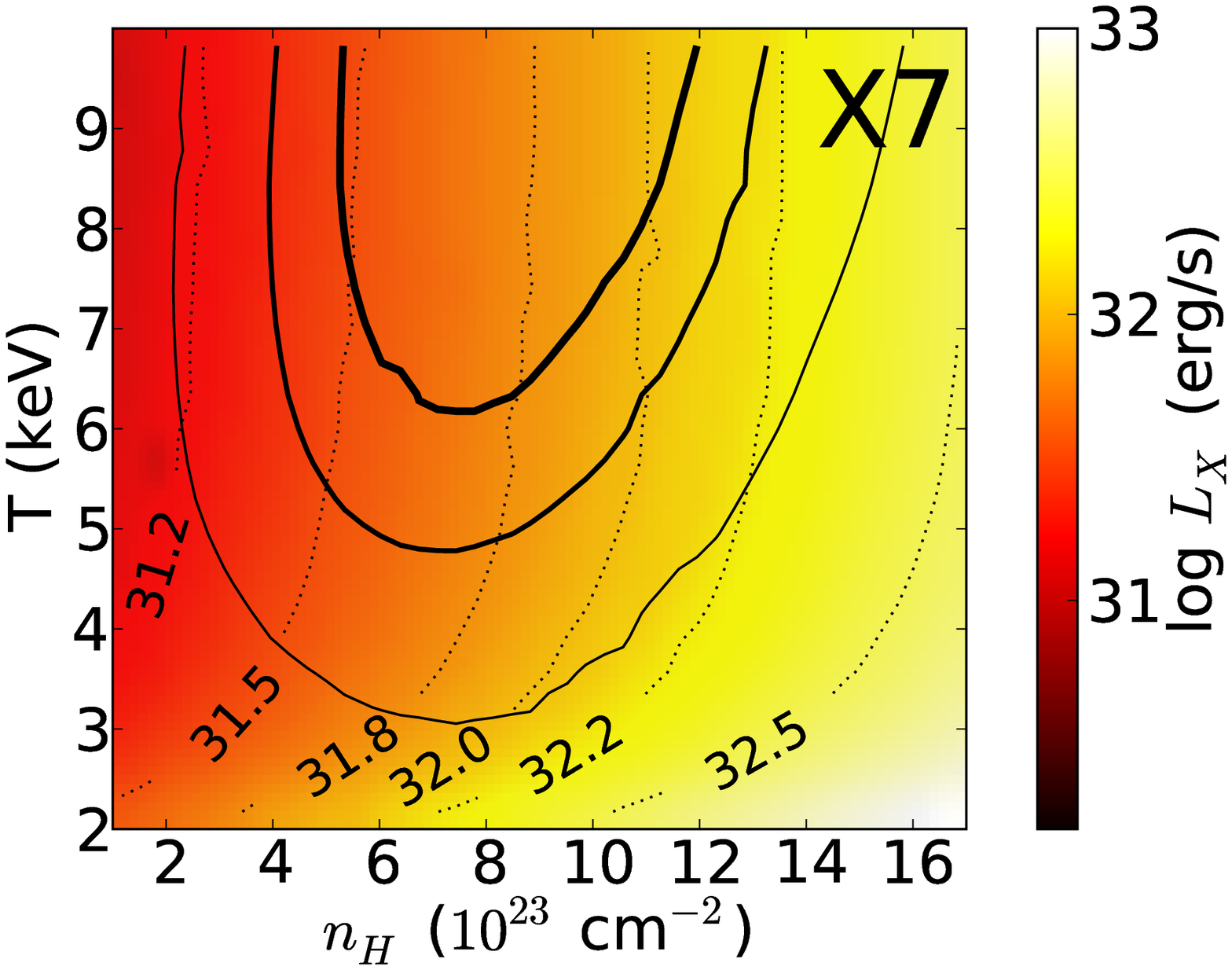}
   \caption{Confidence ranges of the important spectral properties of the four brightest sources. The X-ray luminosity is displayed colorcoded and as dotted contours. The range of spectral parameters  differs between the sources.}\label{fig:contours}%
\end{figure}

Due to the strong absorption, only a relatively low number of counts is available for the individual sources. Fig.~\ref{fig:contours} illustrates the uncertainties of the basic plasma parameters derived for the four most luminous X-ray sources. From this figure it is clear that a more detailed spectral analysis of sources with even fewer counts does not lead to additional insights, since the errors in the parameters already span about an order of magnitude.
The spectra of all sources can be characterised by strong absorption. Virtually any low energy X-ray emission is absorbed, except for X1. This source is absorbed by a column density about an order of magnitude lower than for all other X-ray sources.
In the case of stars, the detected X-ray emission represents the high energy component of the plasma.  Those sources (X3, X6 and X7) with the highest median photon energies do not sufficiently restrict the plasma temperature. Best fit temperatures well above 10~keV are unlikely for the dominating plasma component of (quiescent) stellar sources. Plasma temperatures above 10~keV are rare even for large flares. Therefore, an estimate for the temperature of the hot plasma component is needed, whenever a reasonable fit value is not available. The high temperature component of B-type stars is usually around 2~keV \citep{Stelzer_2005}. We use this as an estimate of the plasma temperature. For comparison we also give the values for $kT = 1.0$~keV. The fits of those sources with a relatively high count number show similar values.
Both temperatures are commonly found for pre-main sequence stars and therefore can be used as an estimate of the plasma temperature almost independently of spectral type.
The unabsorbed luminosity depends strongly on the adopted absorption (see Fig.~\ref{fig:contours} for examples). We therefore give only the range of luminosities corresponding to the lowest and the highest absorption columns~(1$\sigma$). For those sources without a reasonable temperature value, a temperature of $kT=2$~keV was assumed and used to estimate the absorption and the unabsorbed X-ray luminosity; the derived values for $kT=1$~keV are given in brackets.

The lightcurves of the individual objects (Fig.~\ref{fig:lcs}) show some variability, but no flare can be clearly recognised. Usually, flares are also associated with an increase in the median photon energy. Due to the lack of any detected soft emission and the low number of counts, such a signature is not significantly seen except, possibly, in the source X4, which show a monotonic decrease in X-ray luminosity resembling the decay phase of a large flare. During the first half of the observation the photons exhibit a higher value of the mean energy ($kT\approx 4.4$~keV, 46~photons) than during the second half ($kT\approx 4.0$~keV, 18~photons).

Upper limits on the X-ray flux of the non-detected radio sources are estimated by assuming a plasma temperature of $kT=2.0$~keV (1.0~keV) and an absorbing column density of \mbox{$n_H=10^{23}$~cm$^{-2}$}. The maximum source count rate was chosen to be such that for the given limiting X-ray luminosity in 90~\% of the realisations the number of detected photons would be higher than observed. A typical source count number of three photons corresponds to \mbox{ $L_X = 4.8\times10^{29} (2.2\times10^{30}) $~erg$\,$s$^{-1}$}.

There appears to be no correlation between X-ray and radio luminosity for the sources. The X-ray detected radio sources are neither the radio brightest ones nor outstanding in terms of their spectral indices.

\section{Results and individual sources}
We group the sources depending on their association with radio and infrared emission,  starting with the non-detection of the prominent radio source HW~2 (see Fig.~\ref{fig:CenterOverview} or Fig.~\ref{fig:sketch} for a sketch).

\subsection{The non-detection of HW~2\label{sect:HW2-result}}
The central component of the HW~2 emission is not significantly detected in X-rays, only one  photon is observed within a radius of 1~arcsec, where 0.6 photons are expected from the background. Enlarging the extraction region to a radius of 1.5~arcsec also increases the count number to three but overlaps with a source towards the north-east (whose photons are not included). Therefore, the number of source photons is very likely no more than two.
Taking the absorption column density for the region of HW~2 from \citet{Sonnentrucker_2008} of $1.3 \times 10^{23}$~cm$^{-2}$, the upper limit for the X-ray luminosity of HW~2 is $8\times10^{29}$~erg$\,$s$^{-1}$, assuming a plasma temperature of 2.0~keV.  This luminosity is about two orders of magnitude lower than that of a typical mid B-type star since a luminosity of up to $\sim 10^{32}$~erg$\,$s$^{-1}$ is reasonable for massive stars. In this case, the absorbing column density needs to be $>2\times 10^{24}$~cm$^{-2}$ with the other parameters as in the first case to explain the non-detection of HW~2; this value would correspond to $A_V \gtrsim 1000$ according to \citet{Vuong_2003}.
We infer that the absorbing column density towards HW~2 needs to be above a few times $10^{23}$~cm$^{-2}$ to hide any X-ray emission from the presumed B-type star with a 2~keV plasma component.

\begin{figure}
  \centering
   \includegraphics[width=0.24\textwidth]{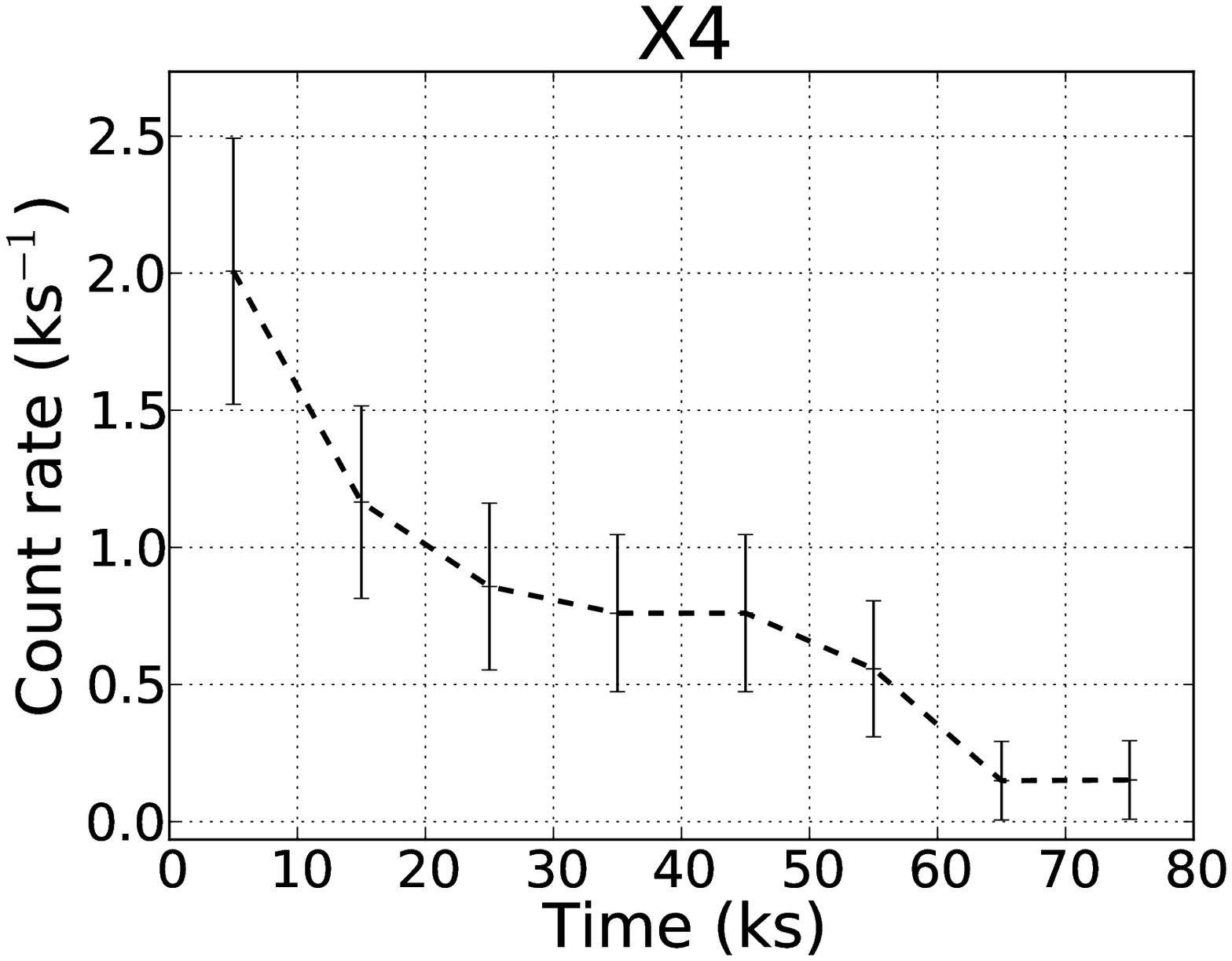}
   \includegraphics[width=0.24\textwidth]{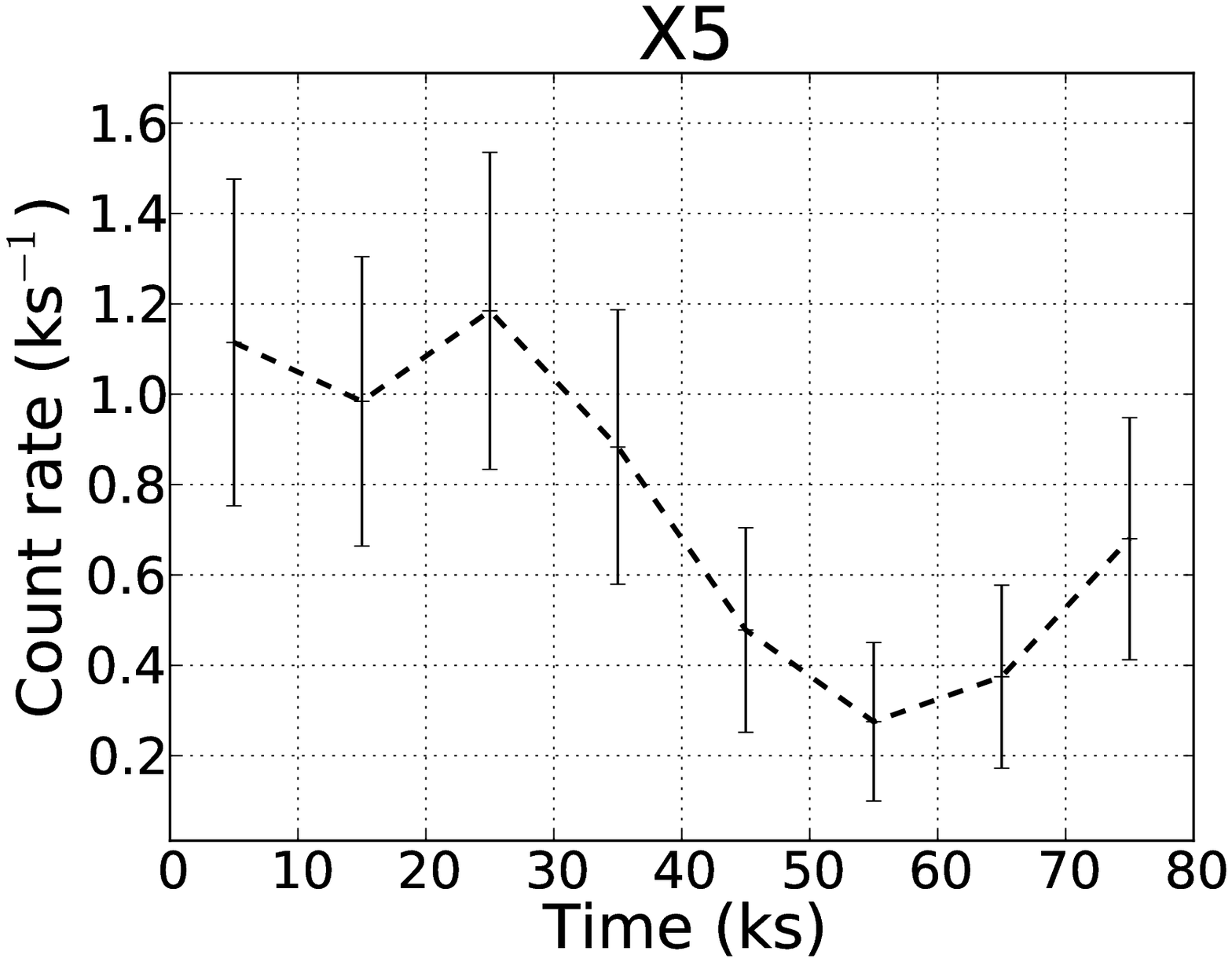}
   \caption{X-ray lightcurves for the sources X4 and X5 (10~ks binning). }\label{fig:lcs}%
\end{figure}

\subsubsection{The HW~2 jet in X-rays?\label{sect:HW2-jet}}
The proper motion of the resolved radio jet  of HW~2 is about 500~km$\,$s$^{-1}$ \citep{Curiel_2006} and therefore sufficient to produce X-rays, when ramming into a medium at rest. Interpolating the apparently linear motion of the radio emitting knots to the date of the {\it Chandra} observation (year 2008), we expect the two outer knots at a separation  of 2.0~arcsec south-west and 1.8~arcsec north-east from HW~2, respectively (see Fig.~\ref{fig:sketch}).
Unfortunately, the radio source HW~8 is located exactly where the south-west knot is predicted. Although no definite explanation for the nature of this highly variable radio source is known, it is likely a pre-main sequence star and also the emitter of the weak X-ray emission at this position. This interpretation is supported by the high median energy of the excess emission. A plasma produced by a shock speed of 500~km$\,$s$^{-1}$ needs to be very luminous to account for the observed hard emission, because the peak of the emission is absorbed and only the highest energy tail remains observable.

At the position of the north-eastern X-ray emission neither radio nor infrared sources are known. The seven photons at exactly the expected position of the radio knot are softer than the south-western ones (median energies of 4.4~keV and 6.2~keV).
Assuming that (at least) the north-eastern X-rays are indeed produced by shocks, what does this imply for the shock luminosity? The velocity of the radio components (500~km$\,$s$^{-1}$) gives a plasma temperature of $kT = 0.3$~keV using the formula of \citet{Raga_2002}. The absorption is unknown, but assuming a value of $n_H = 4\times10^{23}$~cm$^{-2}$, which is on the same order as the value derived by \citet{Sonnentrucker_2008} and compatible with the value found for HW~3c, implies a luminosity of $10^{35}$~erg$\,$s$^{-1}$. Such a high luminosity would require a very efficient process transforming kinetic energy into X-ray emission. For a plasma temperature, on the other hand, of 1.2~keV, the luminosity drops by four orders of magnitude, but requires a high velocity component ($v\sim 1000$~km$\,$s$^{-1}$), embedded in the material observed at radio wavelengths. Such an onion-like structure has been seen in \object{DG~Tau} \citep{Bacciotti_2002}. Using the \citet{Guenther_2009} formula  for the mass loss $\dot{M}$ (eqn.~6), we find  $\dot{M}\approx 2.5\times10^{-5} M_\odot\,$yr$^{-1}$ for the lower temperature and five orders of magnitude smaller for the higher temperature. While the first value is probably too high, the lower value can be easily achieved by pre-main sequence stars like HW~2.
The possibility that the source is a positional chance coincidence is below 1\% (see sect.~\ref{sect:proc}).

The relatively low median energy of the weak excess X-ray emission at exactly the position of the north-eastern radio knot points to shock-induced emission, but requires a higher shock-speed than the velocity of the radio knots. Otherwise the required luminosity is implausibly high. Therefore, either these X-ray photons are caused by an unknown embedded pre-main sequence star, located at the opposite position of HW~8 with respect to HW~2, or these photons are indeed caused by the fastest component of the HW~2 jet.

\subsection{Sources associated with radio sources}
The following X-ray sources are associated with radio emission components.
\subsubsection{HW~9 and HW~8}
The X-ray source X5 clearly coincides with  the radio source HW~9. 
The derived absorbing column density is a few times higher than the value derived by \citet{Sonnentrucker_2008}, compatible with circumstellar matter as expected for the early evolutionary stage of this star. The X-ray light curve shows a variation in the count rate of a factor of three (see Fig.~\ref{fig:lcs}). Although the shape of the lightcurve does not resemble that of large flare events, such variations are typical for young active stars. The unabsorbed luminosity of $5\times10^{32}$~erg$\,$s$^{-1}$ is on the high side for massive stars but not implausible.
All these properties support the idea that HW~9 is of stellar origin (probably of spectral type B), which is in line with the expectations from radio observations \citep{Hughes_1995,Garay_1996}.

At the location of HW~8  a clear photon excess is found (see sect.~\ref{sect:HW2-jet}). The nine photons at the expected position have a very high value of the median energy (6.2~keV) and suggest an interpretation in terms of an embedded pre-main sequence star. Its X-ray luminosity would be $10^{30}$~erg$\,$s$^{-1}$ for an absorbing column density of $10^{23}$~cm$^{-2}$ and $kT=2.0$~keV.

\subsubsection{HW~3b/c (X7) \label{sect:X7}}
The identification of the radio counterpart of X7 is not clear, but it is located closest to HW~3c. Therefore, we favour an association with HW~3c, which is also associated with SMA 875$\,\mu$m emission \citep{Brogan_2007}.
From the X-ray point of view  HW~3c can be characterised by an even higher absorbing column density than HW~9 (X5). The unabsorbed luminosity, assuming a plasma temperature of $kT=1.0$~keV, is unreasonably high for a stellar source or a shocked outflow. Consequently, we favour a model with a higher temperature and less absorption and consequently lower X-ray luminosity. No X-ray photon with an energy lower than 5~keV has been detected. The iron line emission complex around 6.7~keV contributes a large fraction of the observed emission indicative of hot plasma, arguing for a thermal emission component rather than a power-law spectrum.
The radio component 3b also shows a weak excess X-ray emission of six photons with a lower value of the median energy than HW~3c.

In particular, a subcomponent of HW~3d was suspected to be a protostar, driving a jet with a position angle of about 100$^\circ$ \citep{Garay_1996, Torrelles_1998, Goetz_1998}, and therefore a candidate for driving the large scale east-west outflow. This source is not detected in X-rays, possibly HW~3d constitutes only shock induced radio emission without any stellar core.
HW~3a is located further westwards than the other HW~3 radio sources, and might be associated with an infrared source \citep[GPFW~1a,][]{Goetz_1998} without any significant excess X-ray emission.
Still, the same arguments presented in the discussion of the non-detection of HW~2 (sect.~\ref{sect:HW2-result}) apply here: Low-mass YSOs cannot be detected with the available observation, if the strong absorption towards HW~3c is also present towards the other HW~3 components.

\subsection{Sources with infrared counterparts}
In the field presented in Fig.~\ref{fig:CenterXray} four X-ray sources can be identifies with infrared sources.

\subsubsection{X4}
This source is probably associated with the infrared source GPFW2 \citep{Goetz_1998}, which lies at the tip of the radio complex HW~7. \citet{Rodriguez_2005} analysed the proper motions of the individual knots of the HW~7 radio emission complex and postulated that the driving source is located close to the position of X4. We follow this interpretation and regard the radio emission of the HW~7 emission complex as shock heated material \citep[as also proposed by][]{Garay_1996,Goetz_1998}.
The X-ray source itself is absorbed less than those sources close to HW~2 and HW~3. The average luminosity of $L_X \approx 10^{31}$~erg$\,$s$^{-1}$ is very high for a low-mass star, but the lightcurve shows a continuous decrease in count rate during  the observation (Fig.~\ref{fig:lcs}). We interpret it as the decay phase of a strong flare, supported by the lower mean energy of the photons in the second part of the observation (see sect.~\ref{sect:centralregion}). Therefore, the quiescent luminosity of this object is probably significantly lower than the value derived here, thus also a low-mass star can produce the observed X-ray emission.

\subsubsection{X1, X2 and X3}
The X-ray source X1 exhibits the lowest values for the median energy and  the fitted absorbing column density ($n_H = 1.6\times10^{22}$~cm$^{-2}$). Its X-ray luminosity of $6\times10^{30}$~erg$\,$s$^{-1}$ is relatively low in this sample,  but still about ten times higher than the median luminosity of the sources in the {\it Chandra} Orion Ultra Deep project, where a star forming region (Trapezium region) in the Orion Nebula cloud was observed for 730~ks \citep[COUP][]{Getman_2005}; only a tenth of the COUP sources exceed this value.
X1 is also visible in the infrared images of \citet{Lenzen_1984} as IRS~6b.
These properties point to an object located on the near side of the cloud. 

The source X2 is located in the direction of the north-eastern outflow of HW~2 and probably the infrared source IRS~6d.   It is again deeply embedded, and the lower limit of the  absorption is about $7\times10^{22}$~cm$^{-2}$, which matches the expected value at that position. Its high X-ray luminosity ($L_X \gtrsim 3\times10^{30}$~erg$\,$s$^{-1}$) again points to an embedded massive protostar, the coincidence with the outflow is probably a projection effect.

X3 can be characterised by an absorbing column density on the order of a few $10^{22}$~cm$^{-2}$ and a luminosity of $8\times10^{30}$~erg$\,$s$^{-1}$ for a temperature of 2~keV. It is again, most probably, a massive pre-main sequence star. In Fig.~4 of \citet{Goetz_1998} this source seems to be also present as an infrared source, but is not noted as a source by these authors.

\begin{figure}
  \centering
   \includegraphics[width=0.49\textwidth]{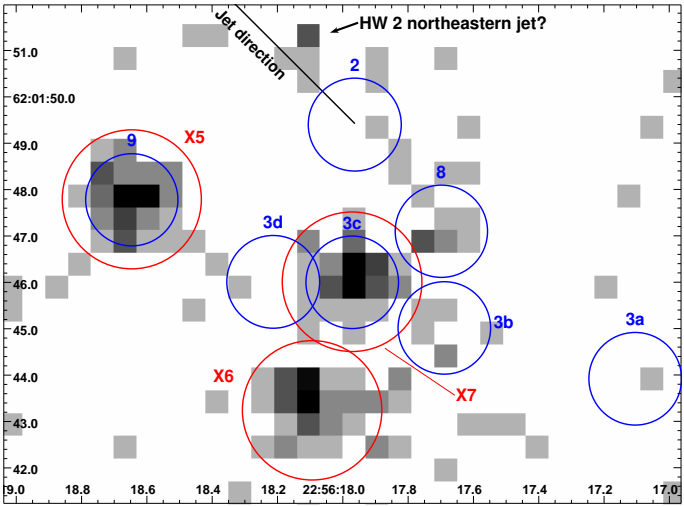}
   \caption{Zoom into Fig.~\ref{fig:CenterXray} for the region around HW~2 and HW~3. Large circles indicate X-ray sources and small circles are radio sources. }\label{fig:CenterCloseup}%
\end{figure}

\subsection{Source without any counterpart (X6)}
The most interesting source in the region unrelated to any radio or infrared source is X6. It is located 2.5~arcsec south of the HW~3 radio emission complex or about 5~arcsec south of HW~2. With a median energy of 5.5~keV, its X-ray spectrum is very hard pointing to a deeply embedded source. The minimum absorbing column density required to explain the spectrum is $3\times10^{23}$~cm$^{-2}$ for a fixed temperature of $2.0$~keV.
This source appears to be slightly extended towards the west, i.e., into the direction of HH~168, where an excess of about 3 photons is seen (see Fig.~\ref{fig:CenterCloseup}).
The available data is not sufficient to exclude the presence of a second object close to the bulk emission causing the distorted structure of this source.
Due to the low count number source variability cannot be excluded. This sources lies close to the connecting line of an H$_2$ emission complex in the west associated with \object{HH~168} and an H$_2$ complex at the eastern side. \citet{Cunningham_2009} estimated the distance of the connecting line to be 5--10~arcsec south of HW~2, which is exactly the value of X6. Thus, X6 is possibly driving the HH~168 outflow.

\subsection{Non-detection of the other radio sources}
From the radio sources which were suspected to harbour an internal power source \citep{Garay_1996}, HW~2, 3a and 3d do not show up in X-rays. The non-detection HW~2 can be easily attributed to the extensive absorbing column density, and HW~3a might simply be too dim for a detection with the available data. It was suggested that this source is a low-luminosity star and thus probably does not reach an X-ray luminosity of a few $10^{29}$~erg$\,$s$^{-1}$ as required for a detection; on the other hand, the radio flux density is higher than in HW~3c, which shows strong X-ray emission.

As B-type stars might potentially exhibit rather soft X-ray spectra, the \textit{Chandra} observation cannot exclude the presence of such stars, since the required X-ray luminosity to shine through an absorbing column density of $10^{23}$~cm$^{-2}$ for a temperature of only 0.3~keV is \mbox{$3\times10^{32}$~erg$\,$s$^{-1}$} which might not be reached by the B-type star. Therefore, in the case of HW~3d, a B-type star can be related to the observed radio emission, although it is not detected in X-rays.

All other radio sources, in particular the HW~7 complex and HW~4, 5, 6 and 1a/b have been interpreted as shock emission \citep{Garay_1996}. The non-detection in X-rays is therefore in line with the expectation for these radio sources.

\section{Summary and conclusion}
Our analysis of the high resolution X-ray observation of the central region of Cep~A relates the detected X-ray sources with known radio and infrared sources of that region. Prior to the \textit{Chandra} observation only very few radio sources had possible counterparts at other wavelengths.  X-rays are thus the 
second energy regime in which more of these sources are detected.

\begin{figure}
  \centering
  \fbox{
    \includegraphics[width=0.49\textwidth]{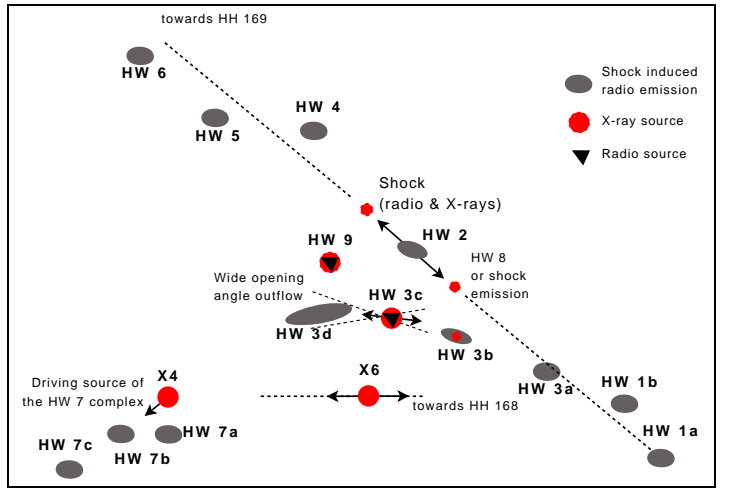}
  }
   \caption{Sketch of a possible scenario. Stellar sources are seen in X-ray and radio emission. The outer jets are mostly found in radio observations, while X-rays originate only close to the jet driving source.}\label{fig:sketch}%
\end{figure}

In Fig.~\ref{fig:sketch} a sketch of a possible scenario including X-ray and radio sources in the central region of Cep~A is shown. This scenario differs in some respects from the \citet{Goetz_1998} picture, but follows the main ideas presented there.
The driving source for HH~169 and HH~174, located in the eastern part of Cep~A, remains the precessing jet of HW~2 \citep{Cunningham_2009}. Although HW~2 is undetected in X-rays, X-ray emission at the expected position of the radio knot within the HW~2 jet is compatible with this scenario. Unfortunately, HW~8 coincides with the expected position of the counter jet and hence prevents us from drawing any conclusion on the origin of the X-ray emission observed at that position. Also, none of the suspected sources in the vicinity of HW~2 are detected, which can be explained by a very high absorbing column density, but requires that \textit{all} those sources are deeply embedded.
IRS~6d is located approximately in the direction of the outflow and also shows X-ray emission (X2), but is probably unrelated to the outflow as its shape is consistent with a point-like source. Furthermore, the non-detection of any of the radio sources, which are located in the direction of the HW~2 jet (HW~1, 4--6), supports the shock interpretation of these radio sources.

The south-eastern radio complex HW~7 has been also interpreted in terms of shock action, supported by the detection of IR~line emission at the tip of this emission complex. The individual components show proper motions pointing to a driving source close to HW~7a \citep{Rodriguez_2005}, which we identify with X4 (GPFW~2), while \citet{Goetz_1998} proposed HW~3d as the driving source. The X4~lightcurve resembles the decay phase of typical flares of active pre-main sequence stars (Fig.~\ref{fig:lcs}), which are known to eject powerful outflows.

For the westward directed outflow (HH~168), another driving source than HW~2 is probably needed, since the position angle of HH~168 with respect to HW~2 differs significantly from that currently observed for its jet. The desired driving source might be HW~3c, the newly discovered X-ray source X6, or a combination of both.
The source of this outflow was not discussed by \citet{Goetz_1998}, but \citet{Cunningham_2009} state that HW~3c is a good candidate for the driving source as also a blue shifted eastward directed CO~lobe emerges from the location of HW~3c.
We speculate that the absence of X-ray emission from the radio complex HW~3d could be explained by interpreting these clumps (also located east of HW~3c) as dense condensations, heated by the outflow of HW~3c (thermal jet emission), while the weak X-ray emission at the position of HW~3b does indeed represent a star.
An opening angle of approximately 20$^\circ$ (position angle of  90$^\circ$) suffices to excite the individual clumps of the HW~3d complex.  This model would be in line with the interpretation of \citet{Brogan_2007}, who detected submillimetre emission towards HW~3c, but not towards HW~3b or 3d. However, the origin of the water maser emission within the HW~3d complex remains unclear in this scenario.
The X-ray source X6 has no counterpart at any wavelength and seems to be elongated towards the west. As it lies closer to the connecting line of HH~168 and its counter outflow, it could also be the driving source.
If the opening of the HW~3c outflow is indeed as large as 20$^\circ$, it is unlikely o drive the large scale outflow, which is more collimated, and X6 would be the natural candidate for driving HH~168.

In summary, the \textit{Chandra} observation of Cep~A detected two of three potential driving sources in the region. HW~3c and the driving source of the HW~7 complex (X4) show the characteristic X-ray properties of pre-main sequence stars. The source of the massive north-east outflow (HW~2) cannot be detected in X-rays, which is explained by the strong absorption towards that position. The X-ray source X6 provides a new candidate for the driving source of HH~168, for which observations at other wavelengths are highly desirable.

{\it Note added in proof: After acceptance of this article \citet{Pravdo_2009} published an analysis of the same X-ray data on astro-ph. Their results concerning the power sources of Cep~A are compatible with ours, they also report the non-detection of HW~2, find the new X-ray source X7 (their source h10), and infer X-ray luminosities of $\log L_X \gtrsim 31$ and plasma temperature above 1~keV for the central sources (see our Tab.~\ref{tab:X-ray}).}

\begin{acknowledgements}
      This work has made use of data obtained from the $Chandra$ data archive.
      P.C.S. acknowledges support from the DLR under grant 50OR0703.
      H.M.G. acknowledges support from the DLR under grant 500R0105.
\end{acknowledgements}
\vspace*{-2mm}
\bibliographystyle{aa}
\vspace*{-4.5mm}
\bibliography{cepA_main}
\end{document}